\documentclass[useAMS,usenatbib,twocolumn]{mn2e}
\usepackage{natbib,amsbsy}
\usepackage{graphicx}
\usepackage{times}
\usepackage{rotate}
\usepackage{color,url}
\begin{document}
\title[Testing redMaPPer Centring Probabilities]
{
Testing redMaPPer Centring Probabilities using Galaxy Clustering and 
Galaxy-Galaxy Lensing
}

\author[Hikage]{Chiaki Hikage$^1$, Rachel Mandelbaum$^2$, Alexie Leauthaud$^{3,1}$,  Eduardo Rozo$^4$, Eli S. Rykoff$^{5,6}$ \\
$^1$ Kavli Institute for the Physics and Mathematics of the Universe (Kavli IPMU, WPI), University of Tokyo, 5-1-5 Kashiwanoha, Kashiwa, Chiba, 277-8583, Japan\\
$^2$ McWilliams Center for Cosmology, Department of Physics, Carnegie Mellon University, Pittsburgh, PA 15213, USA \\
$^3$ Department of Astronomy and Astrophysics, University of California Santa Cruz, Santa Cruz, CA 95064, USA \\
$^4$ Department of Physics, University of Arizona, Tucson, AZ 85721, USA \\
$^5$ Kavli Institute for Particle Astrophysics \& Cosmology, P. O. Box 2450, Stanford University, Stanford, CA 94305, USA \\
$^6$ SLAC National Accelerator Laboratory, Menlo Park, CA 94025, USA
}
\maketitle

\label{firstpage}

\begin{abstract}
Galaxy cluster centring is a key issue for precision cosmology studies
using galaxy surveys. Mis-identification of central galaxies causes
systematics in various studies such as cluster lensing, satellite
kinematics, and galaxy clustering. The red-sequence Matched-filter
Probabilistic Percolation (redMaPPer) estimates the probability that
each member galaxy is central from photometric information rather than
specifying one central galaxy. The redMaPPer estimates can be used for
calibrating the off-centring effect, however, the centring algorithm
has not previously been well-tested. We test the centring
probabilities of redMaPPer cluster catalog using the projected cross
correlation between redMaPPer clusters with photometric red galaxies
and galaxy-galaxy lensing. We focus on the subsample of redMaPPer
clusters in which the redMaPPer central galaxies (RMCGs) are not the
brightest member galaxies (BMEM) and both of them have spectroscopic
redshift. This subsample represents nearly 10\% of the whole cluster
sample.  We find a clear difference in the cross-correlation
measurements between RMCGs and BMEMs, and the estimated centring
probability is $74\pm 10$\% for RMCGs and $13\pm 4$\% for BMEMs in the
Gaussian offset model and $78\pm 9$\% for RMCGs and $5\pm 5$\% for
BMEMs in the NFW offset model. These values are in agreement with the
centring probability values reported by redMaPPer (75\% for RMCG and
10\% for BMEMs) within $1\sigma$. Our analysis provides a strong
consistency test of the redMaPPer centring probabilities. Our results
suggest that redMaPPer centring probabilities are reliably
estimated. We confirm that the brightest galaxy in the cluster is not
always the central galaxy as has been shown in previous works.
\end{abstract}

\begin{keywords}
galaxies: clusters: general
\end{keywords}

\section{Introduction}
In the standard hierarchical structure formation scenario, small dark
matter haloes form first and grow by accretion of surrounding matter
and also by merging other haloes. When a small halo merges into a more
massive halo, the galaxy that was originally the central galaxy in the
small halo becomes a satellite galaxy in the merged halo, while the
central galaxy hosted by the more massive halo remains the central
galaxy of the final dark matter halo.  After merging, the star
formation of satellite galaxies is quenched. The central galaxy
continues to accrete new gas from satellite galaxies. In consequence,
the central galaxy is luminous, massive, and is located near the
bottom of the halo potential well.  Actually central galaxies form
distinct population from other smaller neighboring galaxies, i.e.,
satellites, in many aspects such as color, star formation activity,
AGN activity, morphology, and stellar populations
\citep{Weinmann06,vonderLinden07,vandenBosch08,Skibba09,Hansen09}.

We define the central galaxy as the one with the lowest
  specific potential energy in each cluster. Identifying central
galaxies in galaxy groups and clusters is important for both cosmology
and galaxy evolution studies.  Off-centred clusters that are used in
stacked lensing analysis without any modeling to account for the
off-centring cause systematics in the mass estimation
\citep{Johnston07,Leauthaud10,Okabe10,OguriTakada11,George12,Hikage13}.
The clustering properties of central and satellite galaxies are
different and this features in the commonly used halo occupation
modeling
\citep[HOD:][]{Seljak00,Scoccimarro01,BerlindWeinberg02,Zheng05,Masjedi06,ReidSpergel09,White11,Leauthaud11,Coupon12,Manera13,Parejko13}.
Internal motions of satellite galaxies generate the non-linear
redshift-space distortion, which is a systematic uncertainty in
cosmology studies from redshift-space galaxy clustering
\citep{CabreGaztanaga09,Reid09,Samushia09,Hikage12,HikageYamamoto13,Guo15}.
In the studies of satellite kinematics, off-centred galaxies lead to
overestimate halo masses \citep{Skibba11}.

Identifying central galaxies are not always simple unless there is a
dominant cD galaxy at the X-ray peak emission. The central galaxy is
often assumed to be the brightest galaxy in the cluster (BCG)
\citep{vandenBosch04,Weinmann06,Budzynski12}. There are, however, many
studies indicating that a significant fraction of central galaxies are
not BCGs
\citep{vandenBosch05,vonderLinden07,Coziol09,Sanderson09,Einasto11,Skibba11,George12,Hikage13,Sehgal13,Lauer14,Hoshino15}.
For example, \cite{Skibba11} use phase-space statistics to find that
the off-centring fraction, i.e., the fraction of BCGs that are not the
central galaxy, increases from $\sim 25$\% for Milky-Way size halos to
$\sim 45$\% for massive clusters. \cite{Hikage13} use the
redshift-space power spectrum, galaxy-galaxy lensing, and the
cross-correlation with photometric galaxies, and found that the
off-centring fraction for massive clusters is $46\pm
5$\%. \cite{Lauer14} use nearby clusters to find that $\sim 15$\% of
BCGs have large offsets ($>100$kpc) from X-ray centres.  While
high-resolution X-ray and SZ data can help robustly identify the
deepest part of the cluster potential well and therefore the true
cluster central galaxy
\citep[e.g.,][]{Ho09,George12,Stott12,vonderLinden12,Menanteau13,Mahdavi13,RozoRykoff14,Oguri14},
the size of the cluster sample with the necessary data is quite
limited.

The red-sequence Matched-filter Probabilistic Percolation (redMaPPer)
is a red-sequence cluster finding algorithm
\citep{Rykoff14,RozoRykoff14}, which is optimized for multi-band
photometric surveys such as the Sloan Digital Sky Survey
\citep[SDSS;][]{York00}, the Hyper-Suprime Cam \citep[HSC;][]{HSC}
survey, the Dark Energy Survey \citep[DES;][]{DES}, and the Large
Synoptic Survey Telescope \citep[LSST;][]{LSST}. A similar
red-sequence based cluster finding algorithm called CAMIRA (Cluster
finding Algorithm based on Multi-band Identification of Red-sequence
gAlaxies) has also been developed to provide a cluster catalog using
Hyper-Suprime Cam data in a wide range of redshifts
\citep{Oguri14,Oguri17}.  A key feature of the redMaPPer algorithm is
that it estimates the centring probability of every member galaxy
rather than identifying from three photometric observables:
luminosity, color, and the local galaxy density. redMaPPer
  estimates of centring probabilities are useful for calibrating the
  systematics associated with the off-centring effect in various
  cosmology studies such as cluster mass estimates using stacked
  lensing analysis and satellite kinematics and also the accurate
  modeling of the redshift-space clustering. However, there was not
  enough verification of the redMaPPer centring probabilities.
\cite{Hoshino15} found that the off-centring probability of the
brightest cluster member galaxy is 20-30\% using the redMaPPer
centring probability, while \cite{Skibba11} found that the
off-centring value is 40\% using phase-space statistics.  The results
in \cite{Hoshino15} rely on the assumption that the redMaPPer central
probability is accurate. Validating that result and many other
cosmological studies using the redMaPPer algorithm requires us to
verify that the redMaPPer centring probability is an accurate
reflection of reality. The accuracy of the redMaPPer centring
probability was previously studied by comparison with X-ray clusters
\citep{RozoRykoff14}. The overlapped area for the comparison is small
and thus the result depends on the selected X-ray subsamples.

Cluster-galaxy lensing, a cross-correlation of foreground clusters
with background galaxy shapes, provides an estimate of the projected
mass distribution around the foreground clusters
\citep[e.g.,][]{Mandelbaum06,Sheldon09,Leauthaud10,Okabe10,Mandelbaum13}.
When the lens sample contains off-centred (satellite) galaxies, the
stacked lensing signals are modified in a way that depends on the
satellite fraction and radial offsets
\citep{Johnston07,Leauthaud10,Okabe10,OguriTakada11,George12,Hikage13}.
The cross-correlation with photometric galaxies provides another way
to evaluate the central fraction of a given galaxy sample
statistically \citep{Mandelbaum13,Hikage13}.

In this paper, we test the redMaPPer centring probability using the
cross-correlation measurements, and verify the results using
cluster-galaxy lensing.  Here we focus on the clusters for which the
brightest member galaxy does not have the highest central galaxy
probability according to the redMaPPer algorithm. These clusters
usually do not have well-defined central galaxies, and therefore
provide a good sample to check whether the redMaPPer centring
probability agrees with the estimation from the cross-correlation
measurements. We also select a sample with highly secure central
galaxy identification as a reference sample.

This paper is organized as follows: in section 2 we describe the
details of the galaxy cluster sample.  In section 3 we explain how to
estimate cross-correlation with the photometric red galaxies and
galaxy-galaxy lensing. In section 4, the theoretical model to compare
with the observations to estimate the off-centring properties is
described. In section 5, we present the results of the
cross-correlation measurements with different proxies of central
galaxies and constraints on their central fraction.  Section 6 is
devoted to summary and conclusions. Throughout this paper, we adopt a
flat $\Lambda$ cold dark matter cosmology with $\Omega_m=0.3$ and
$\sigma_8=0.8$.  We use physical (not comoving) units for distances
and lensing signals.

\section{Data}
\subsection{Galaxy cluster samples}
redMaPPer is a red-sequence cluster finding algorithm \citep[see the
  details of the algorithms in][]{Rykoff14,RozoRykoff14}. It utilizes
the 5-band ($ugriz$) imaging data to identify galaxy clusters by
characterizing the evolution of red sequence in an iterative
self-training technique. redMaPPer estimates the photometric redshift
of each cluster $z_{\lambda}$ and the cluster richness $\lambda$ from
the membership probability assigned to each galaxy in a cluster field
\citep{Rozo15}. Our sample is constructed from the redMaPPer v5.10
cluster catalog \citep{Rozo15} based on the Sloan Digital Sky Survey
Data Release 8 (SDSS DR8) photometric data \citep{SDSSDR8}. The sample
covers about 10,000 deg$^2$ contiguous sky area with the same survey
mask of Baryon Oscillation Spectroscopic Survey
\citep[BOSS;][]{Dawson13}. We extract a sample of 7730 redMaPPer
galaxy clusters with $\lambda \ge 20$ in the range $0.16< z_\lambda <
0.33$, which is identical to the sample analyzed by \citep{Hoshino15}.
We impose an additional mask for the cross-correlation and lensing
measurements to match the area coverage of the fainter sample used for these cross-correlations
 \citep[see the details in
][]{Reyes12,Nakajima12,Mandelbaum12,Mandelbaum13}

\begin{figure*}
\begin{center}
\includegraphics[width=5.3cm]{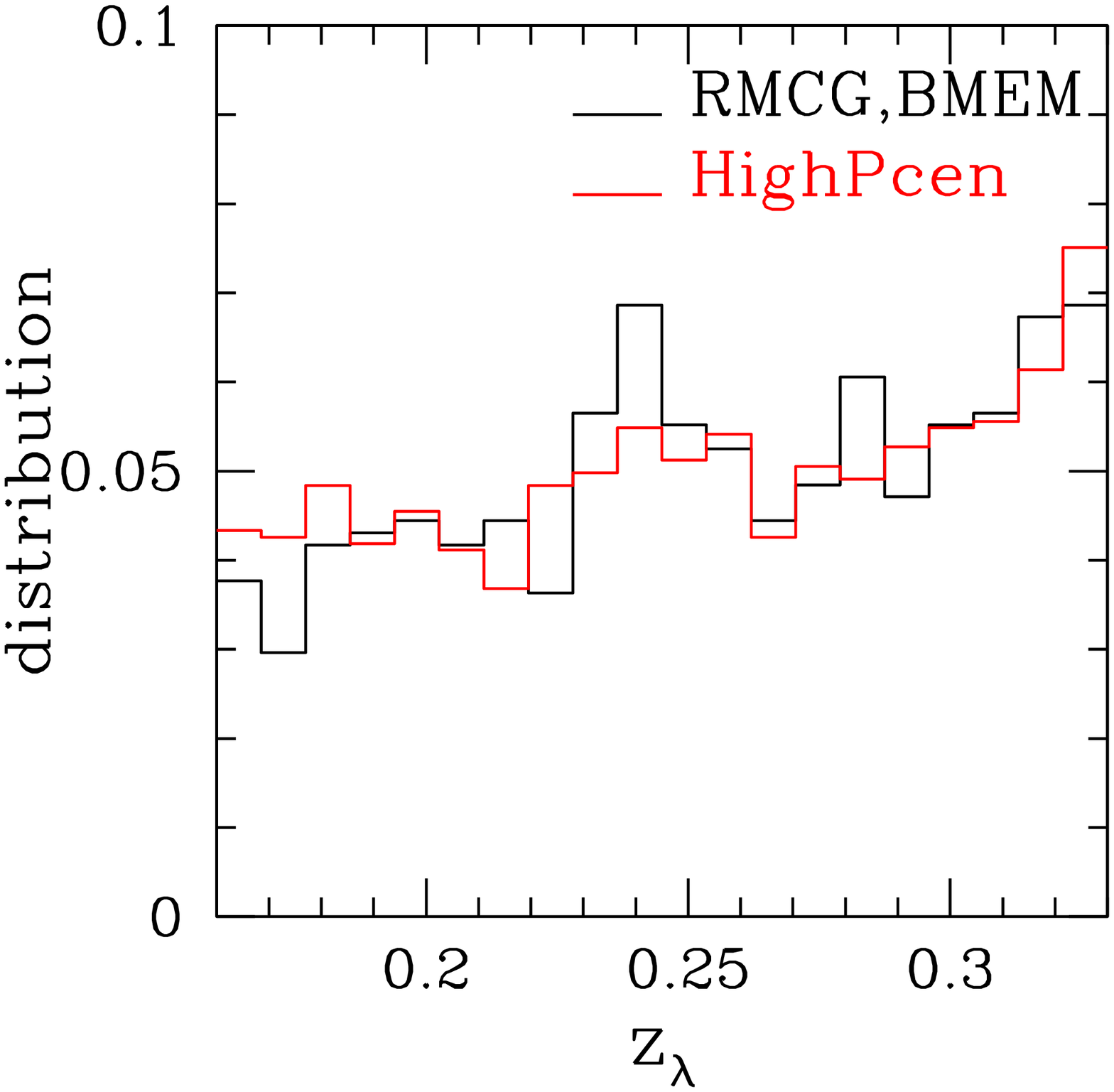}
\includegraphics[width=5.4cm]{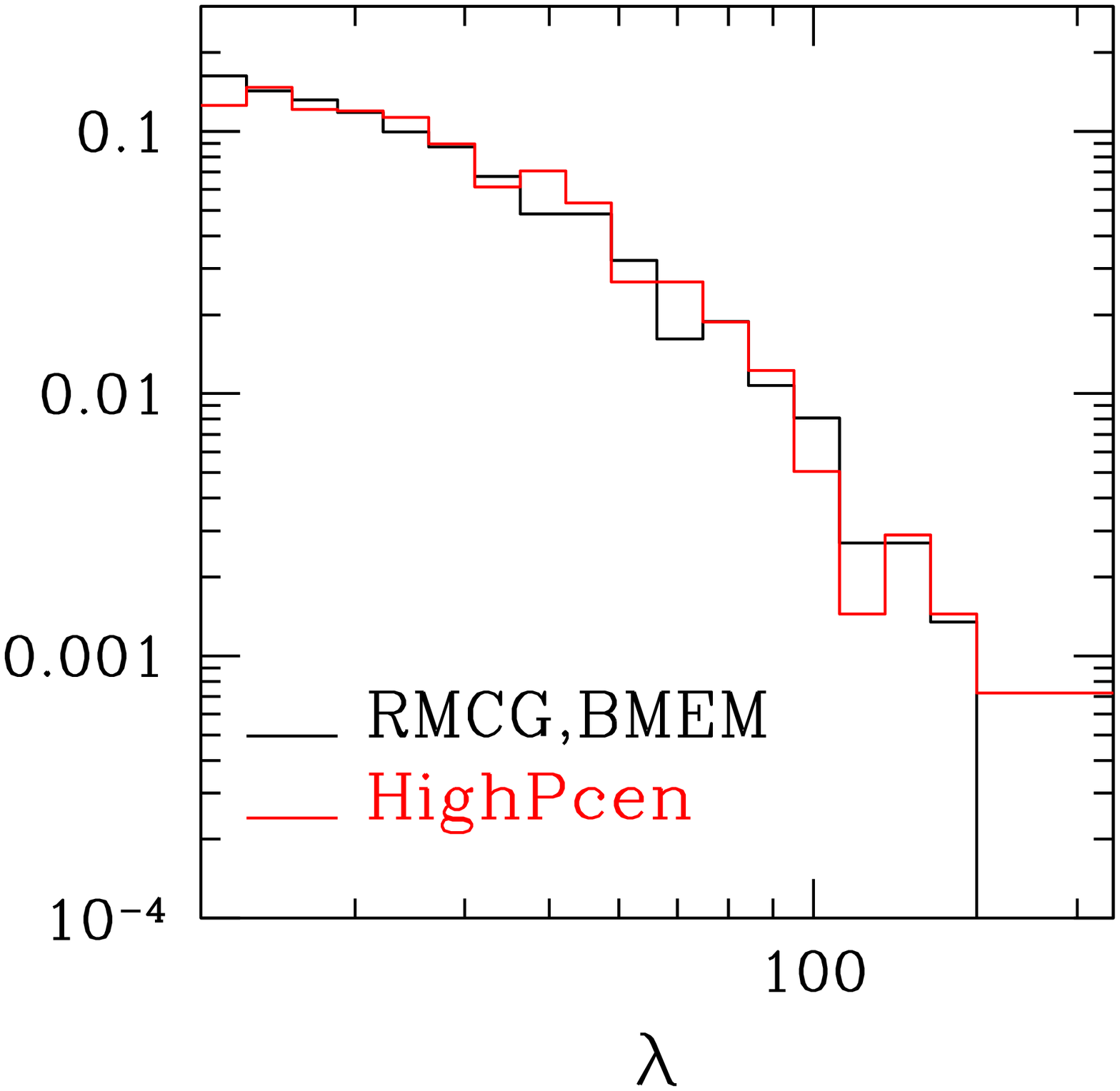}
\includegraphics[width=5.5cm]{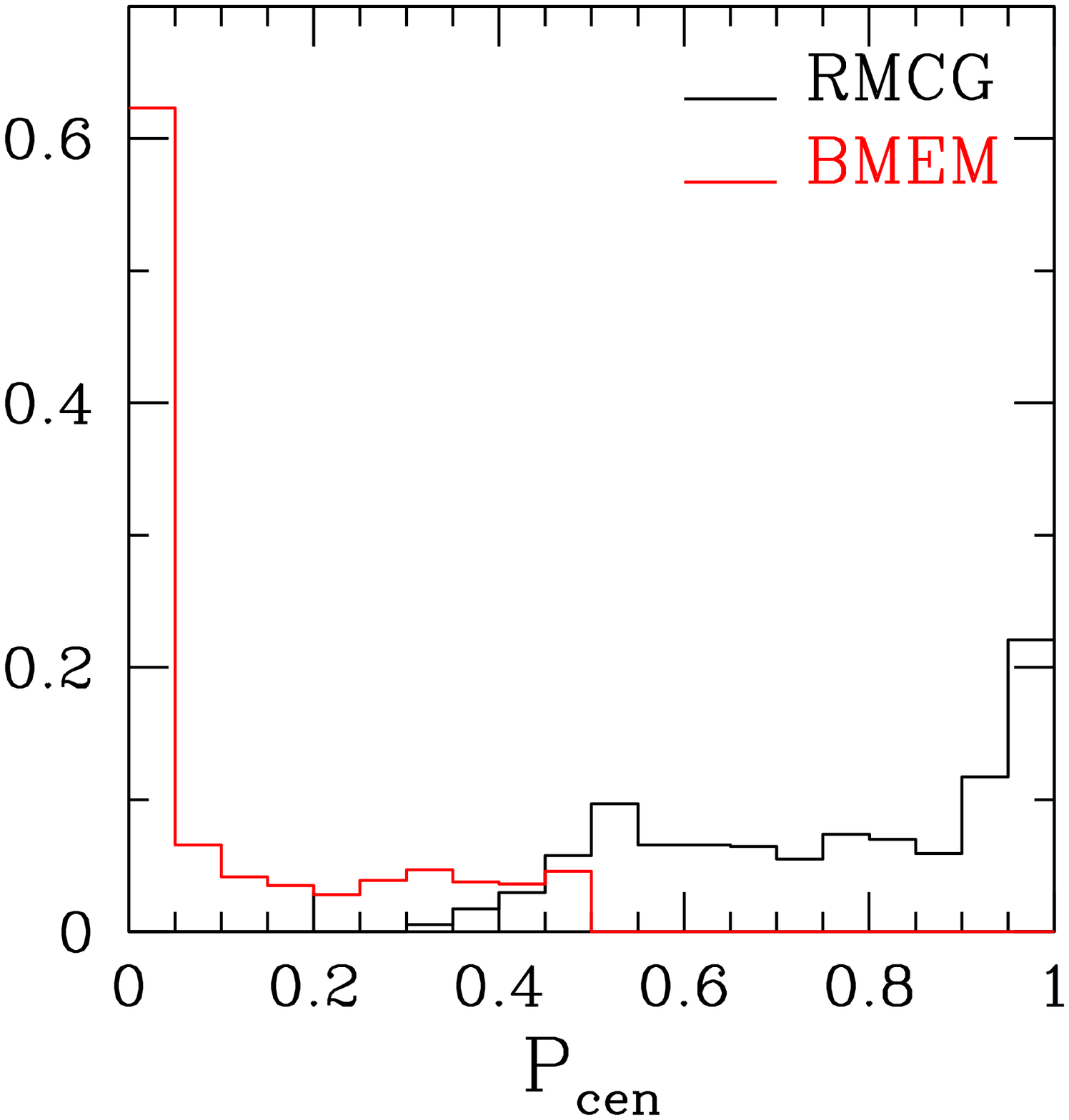}
\end{center}
\caption{Histograms of photometric redshift $z_\lambda$ (left)
    and richness $\lambda$ (middle) for redMaPPer central galaxies
    (RMCG), brightest member galaxies (BMEM) and High Pcen sample,
    which is the reference sample of 'central' galaxies with the
    redMaPPer centring probability higher than 99\%.  The details of
    the samples are written in the text. The distributions of both
    $z_\lambda$ and $\lambda$ are similar among the samples. The right
    panel shows the $P_{\rm cen}$ distribution for the RMCG and BMEM
    samples. The histogram is normalized by the total number of galaxy
    clusters in each sample. Since the RMCG and BMEM samples are
    specifically chosen to include clusters for which the RMCG and
    BMEM galaxy are not the same, the redMaPPer centring probability
    of BMEMs is much lower than that of RMCGs. This is not the case
    when considering all redMaPPer clusters.}
\label{fig:hist}
\end{figure*}

A key feature of redMaPPer is that it assigns a centring probability
$P_{\rm cen}$ to every member galaxy. When there is a well-defined central galaxy, the
$P_{\rm cen}$ of the galaxy approaches unity. Otherwise $P_{\rm cen}$
may be well below unity. The centring probability $P_{\rm cen}$ of
every member galaxy candidate is estimated from three filters:
\begin{equation}
u_{\rm cen}=\phi_{\rm cen}(m_i|z_\lambda,\lambda)G_{\rm cen}(z_{\rm red})
f_{\rm cen}(w|z_\lambda,\lambda),
\end{equation}
where the luminosity filter $\phi_{\rm cen}$ uses the $i$-band
magnitude $m_i$, the photometric redshift (color) filter $G_{\rm cen}$
uses the photometric redshift $z_{\rm red}$, and the local density
filter $f_{\rm cen}$ depends on the gravitational potential weight
$w$. A higher value is returned for galaxies where the luminosity is
higher in $\phi_{\rm cen}$, the photometric redshift is closer to the
cluster redshift in $G_{\rm cen}$, and the local density is higher in
$f_{\rm cen}$. Each filter is assumed to be a Gaussian distribution
with both its mean and its dispersion in $\phi_{\rm cen}$ and $f_{\rm
  cen}$ dependent on $z_{\rm \lambda}$ and $\lambda$. The dispersion
of $z_{\rm red}$ in $G_{\rm cen}$ is set to be broad so that the
galaxies that have slightly offset colors are allowed to be central
galaxies. Since the redMaPPer centring probability depends on the
above three observables, the brightest member galaxy (hereafter
denoted by BMEM) does not always have the highest centring probability
when the galaxy is isolated from other galaxies. In this paper, we
call the galaxy with the highest centring probability ``redMaPPer
central galaxy'' (hereafter denoted by RMCG).

We focus on clusters for which the RMCG and BMEM are not the
same. Since these clusters usually lack a well-defined central galaxy,
they are particularly useful for testing whether the redMaPPer
centring probability is correct. An example of such clusters is shown
in Fig.6 of our previous paper \citep{Hoshino15}.  In order to reduce
contamination from projection effects, we further restrict the sample
to the clusters for which both the RMCGs and BMEMs have confirmed
spectroscopic redshifts in the BOSS LOWZ DR12 sample \citep{SDSSDR12}
and their redshifts are consistent within $3\sigma$, i.e., $c|z^{\rm
  RMCG}-z^{\rm BMEM}|/(1+z^{\rm RMCG})<3\sigma_v$, where $c$ is the
speed of light and $\sigma_v$ is the velocity dispersion inside
clusters. We adopt a richness- and redshift-dependent $\sigma_v$ in
the form of
$\sigma_v=\sigma_p((1+z_\lambda)/(1+z_p))^\beta(\lambda/\lambda_p)^\alpha$
with the best-fitting value of $\sigma_p=618.1$~km/s, $z_p=0.171,
\lambda_p=33.336, \alpha=0.435, \beta=0.54$ fitted to the redMaPPer
clusters \citep{Rozo15}. The final number of galaxy clusters is
reduced to 743, which is 9.6\% of the total cluster sample in the same
range of $z_\lambda$ and $\lambda$. The average richness and redshift
of this sample is $\langle\lambda\rangle=36.1$ and $\langle
z_\lambda\rangle=0.253$. The mean centring probability values that
are reported by redMaPPer for this sample are 0.754 for RMCGs and
0.103 for BMEMs.
The distributions of $z_\lambda$, $\lambda$ and $P_{\rm cen}$ for
RMCG and BMEM samples are plotted in Figure~\ref{fig:hist}.
The distributions are normalized to the total number of galaxy
clusters in the sample.

As a reference sample of central galaxies, we construct a ``High Pcen'' sample for which
the centring probability of RMCGs is larger than 99\% and for which the
RMCG has a spectroscopic redshift that is consistent with the
spectroscopic redshift of at least one other cluster member. 
The number of galaxies in the ``High Pcen'' sample is 1385 and its average
$\langle \lambda\rangle=37.3$ and $\langle z_\lambda\rangle=0.251$,
consistent with the RMCG/BMEM samples. 
The distributions of $z_\lambda$ and $\lambda$ for the High Pcen sample
are also plotted in Figure~\ref{fig:hist}. 

In summary, we prepare three ``central galaxy'' samples:
\begin{enumerate}
\item{RMCG}: clusters with different RMCG and BMEM, and the centre defined as RMCG
\item{BMEM}: clusters with different RMCG and BMEM, and the centre defined as BMEM
\item{High Pcen}: clusters with the centring probability of RMCG is higher than 99\% 
\end{enumerate}
We remake random catalogs to match the distribution of $\lambda$ and
$z_\lambda$ in each sample, and the same sky coverage as the whole cluster catalog.

\section{Measurements}
\subsection{Projected cross-correlation of galaxy clusters with photometric red galaxies} 
In order to constrain the off-centring fraction in each ``central
galaxy'' sample described in the previous section, we use the projected
cross-correlation function of each central galaxy with a sample of
photometric red galaxies selected from the source catalog from SDSS
DR8. The spectral energy distribution (SED) and photo-$z$ are estimated
with the Zurich Extragalactic Bayesian Redshift Analyzer 
\citep[ZEBRA;][]{Feldmann06}. We use red galaxies by selecting those galaxies for which the ZEBRA
results are consistent with an early-type galaxy SED, and a cut is imposed on the extinction-corrected model
magnitude at $r<21$ 
to restrict to a sample with smaller photometric redshift errors. Systematic tests of the redshifts
for this photometric sample were carried out in \cite{Nakajima12}.

The projected cross-correlation function for each central galaxy sample
(RMCG, BMEM, or High Pcen) is estimated as follows:
\begin{equation}
w^{\rm cross}(R)=\frac{D^{\rm (cluster)}D^{\rm (photo-z)}}{R^{\rm (cluster)}D^{\rm (photo-z)}}-1,
\end{equation}
where $D^{\rm (cluster)}D^{\rm (photo-z)}$ is the number of pairs between
cluster central galaxies and photo-$z$ galaxies at the projected
separation of $R$ (calculated at the cluster redshift).  For this purpose, we
only use pairs for which the photo-$z$ is consistent with the cluster
redshift within $1\sigma$ and the number is normalized with the total
number of pairs. $R^{\rm (cluster)}D^{\rm (photo-z)}$ is same as
$D^{\rm (cluster)}D^{\rm photo-z}$ but for the number of pairs between
a random catalog corresponding to the cluster distributions and photo-$z$ galaxies. The amplitude of the
cross-correlation depends on the photo-$z$ error because the photo-$z$
galaxies that are not physically associated with the galaxy clusters
dilute the cross-correlation signal. As a result, we marginalize over the amplitude information in
our analysis. 

The covariance matrix for the projected clustering signal, and the
cross-covariance between different samples, is estimated using the
jackknife resampling method, which for clustering seems to overestimate the covariance by about 10\% 
on scales below the jackknife region size
\citep{Singh17}.  The scales that dominate our constraints all satisfy
this requirement.

\subsection{Galaxy-galaxy lensing}

We also use the galaxy-galaxy lensing, the cross-correlation of each
central galaxy with shapes of background galaxies.  Since this measurement is directly sensitive to
the matter distribution around the lens galaxies, it is also a 
useful probe of off-centring of these galaxies within their host halos. The galaxy-galaxy lensing
measurement, however, has a larger statistical error compared to the
cross-correlation measurements and is primarily used for a consistency check.

The source galaxy sample is selected from the same source catalog as described
above. The source galaxy shapes are measured with the 
re-Gaussianization method \citep{HirataSeljak03}. We use a shear
catalog with a number density of 1.2 per square arcmin,
which was first presented in \cite{Reyes12} and further tested and validated in later work
\citep{Nakajima12,Mandelbaum12,Mandelbaum13}. For the lensing
measurement, we use source galaxies to the faint limit of the catalog (extinction-corrected model
magnitude $r<21.8$) with photometric redshift
larger than each central galaxy. Galaxy-galaxy lensing is measured
with the differential projected mass density $\Delta\Sigma (R)$, which
is optimally estimated as \citep{Mandelbaum06}
\begin{eqnarray}
\Delta\Sigma(R)&=&\frac{\sum_{\rm ls} w_{\rm ls}\gamma_t^{\rm (ls)}(R)\Sigma_{\rm
    crit}(z_l,z_s)}{\sum_{\rm rs}w_{\rm rs}} \nonumber \\
&& -~\frac{\sum_{\rm rs} w_{\rm rs}\gamma_t^{\rm (rs)}(R)\Sigma_{\rm
    crit}(z_r,z_s)}{\sum_{\rm rs}w_{\rm rs}},
\end{eqnarray}
where $\gamma_t$ is the tangential shear with respect to each lens galaxy
and $\Sigma_{\rm  crit}$ is the critical surface mass density given as
\begin{equation}
\Sigma_{\rm crit}(z_l,z_s)=\frac{c^2}{4\pi G}\frac{D_A(z_s)}{D_A(z_l)D_A(z_l,z_s)},
\end{equation}
with the angular diameter distance $D_A(z)$.  The subtraction of tangential shear around random
points removes contributions from additive systematics and results in a more optimal estimator
\citep{Singh17}.  The division by the sum of weights around random points 
accounts for the dilution of the signal by unlensed cluster member galaxies that are included in the
source sample due to photo-$z$ error \citep{Mandelbaum05}. 
The shear weight for
galaxies and random, $w_{\rm ls}$ and $w_{\rm rs}$ respectively, is
given as the inverse variance weight
\begin{equation}
w=\Sigma_{\rm crit}^{-2}(z_l,z_s)(\sigma_{\rm int}^2+\sigma_{\rm i}^2),
\end{equation}
where $\sigma_{\rm int}$ is the intrinsic ellipticity fixed to be
0.365 and $\sigma_{\rm i}$ is the shape measurement error due to pixel
noise \citep{Reyes12}.

As illustrated in \cite{Singh17}, for g-g lensing, the jackknife method properly represents the
covariance below the jackknife region size.  This validates our choice of jackknife errorbars.

\section{Theoretical modeling}
In this section, we provide a theoretical modeling of the projected
cross-correlation $w(R)$ and galaxy-galaxy lensing $\Delta\Sigma(R)$
based on the previous work by \cite{Hikage13}.

In order to take into account the halo mass dependence of the radial
profile of photo-$z$ galaxies, we convert the richness distribution of
the clusters $P(\lambda)$ to the halo mass distribution $P(M)$ by
\begin{equation}
\label{eq:massdist}
P(M)=\int d\lambda P(\lambda) P(M|\lambda),
\end{equation}
where $P(M|\lambda)$ is the conditional mass distribution for a fixed
$\lambda$. Here we use $P(M|\lambda)$ calibrated by the galaxy-galaxy
lensing analysis of \citet{Simet17}. They model $P(M|\lambda)$ as a
lognormal distribution with a mean mass-richness relation that is
parametrized as
\begin{equation}
\langle M|\lambda \rangle = M_0\left(\frac{\lambda}{\lambda_0}\right)^\alpha ,
\end{equation}
where the pivot scale of richness $\lambda_0$ and $\alpha$ is the
power-law slope of a mass at a given $\lambda$. The scatter in mass at
a fixed richness is given as the sum of a Poisson term and an
intrinsic variance term as
\begin{equation}
{\rm Var}(\ln M|\lambda)=\frac{\alpha^2}{\lambda}+\sigma_{\ln M|\lambda}^2.
\end{equation}
The halo mass $M$ is defined as $M_{200m}$, the mass enclosed in a
sphere of 200 times the mean matter density. We use their best-fit
values $\log(M_0/h^{-1}M_\odot)=14.344$ at $\lambda_0=40$, and
$\alpha=1.33$. The variance $\sigma_{\ln M|\lambda}$ is set to be
0.25. The mean mass $\log(M/h^{-1}M_\odot)$ is 14.3 for both of our
RMCG/BMEM and High Pcen samples.

\begin{figure*}
\begin{center}
\includegraphics[width=16cm]{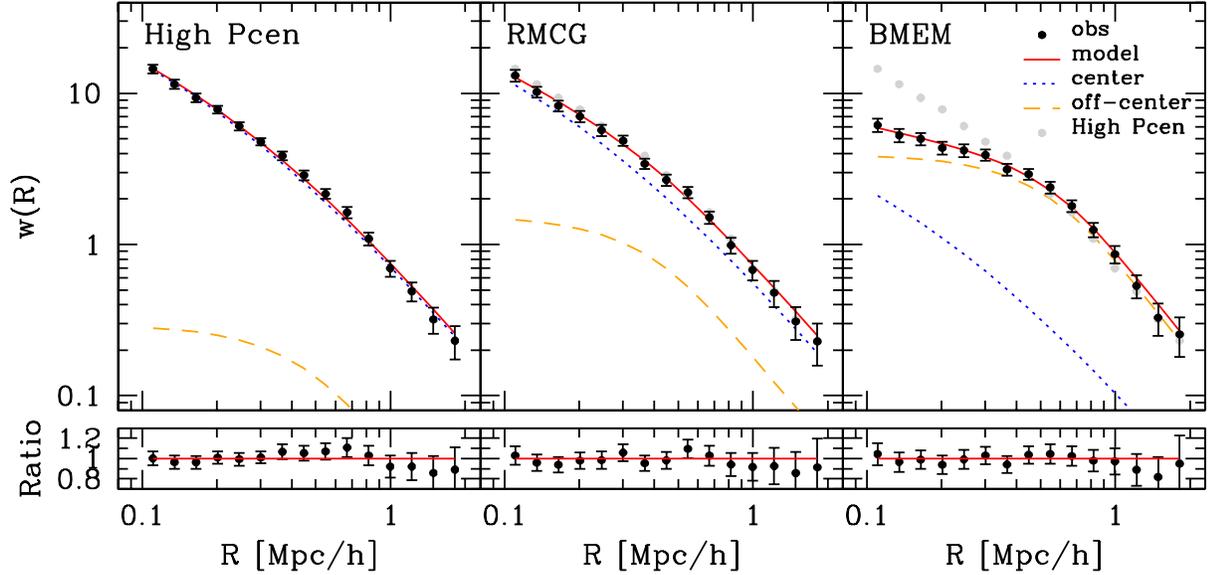}
\end{center}
\caption{Cross-correlation measurements of photometric red galaxies
  with High Pcen galaxies (left), redMaPPer central galaxies (middle)
  and brightest galaxy samples (right) are shown as black filled
  circles.  Data points for the High Pcen sample are plotted in the
  right two panels with thin gray points for reference.  We also plot
  the best-fitting model (Eq.~\ref{eq:wR}, red solid lines); centring
  and off-centring components are plotted with blue dotted and yellow
  dashed lines respectively. The error is estimated from jackknife
  resampling. As shown, the clustering of red galaxies around
    RMCGs is close to be that of red galaxies around High Pcen sample,
    and the profile is much steeper than that of red galaxies around
    BMEMs. This indicates that the centring fraction of RMCGs is much
    larger than BMEMs.  Lower panels show the ratio of observed
  values to the best-fitting model.}
\label{fig:fit_angcor}
\end{figure*}
\begin{figure*}
\begin{center}
\includegraphics[width=16cm]{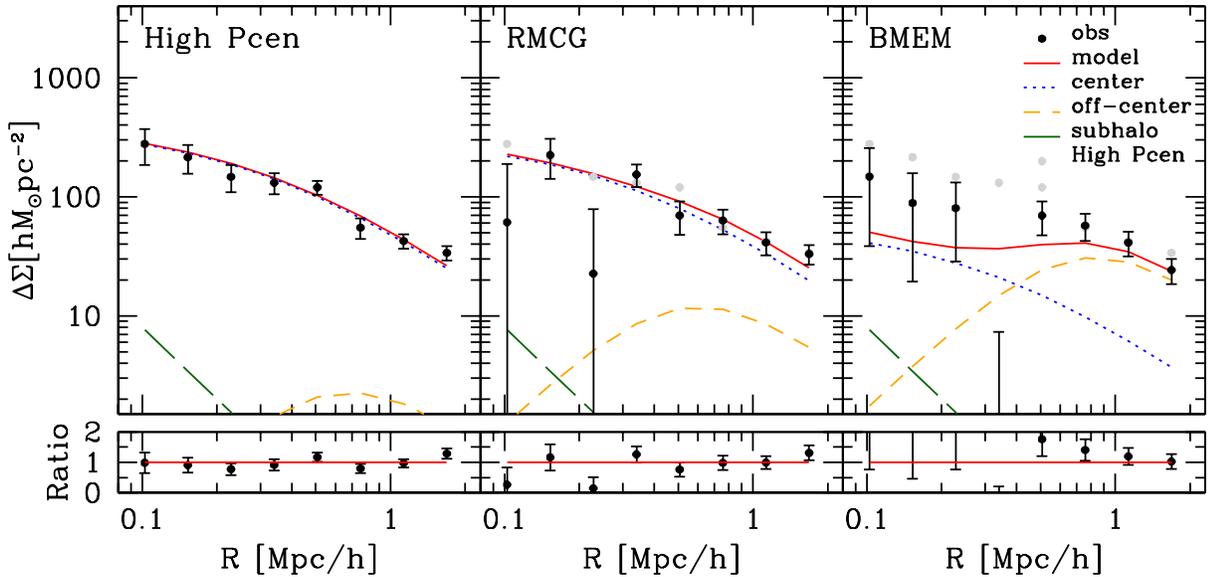}
\end{center}
\caption{Same as Fig.~\ref{fig:fit_angcor} but for the galaxy-galaxy
  lensing $\Delta\Sigma$. We use the model in
  Equation~(\ref{eq:dsigma}) with the best-fitting values of $q_{\rm
    cen}$ and $\tau_{\rm off}$ obtained from the cross-correlation
  $w(R)$. We set $\log(M_*/h^{-1}M_\odot)=11.4$. While the
    lensing data are too noisy to fit a new model, this figure
    demonstrates that the lensing data are consistent with the model
    derived from the cross-correlation measurements.}
\label{fig:fit_dsigma}
\end{figure*}

We employ the Navarro-Frenk-White (NFW) profile \citep{NFW} to
describe the distribution of the red galaxies inside halos 
by replacing the mass with the galaxy number:
\begin{equation}
\rho_g(r)=\frac{\rho_{g,s}}{(r/r_s)(1+r/r_s)^2},
\end{equation}
where the density parameter $\rho_{g,s}$ is given by
\begin{equation}
\rho_{g,s}=\frac{N_g(M)}{4\pi r_s^3(\log(1+c)-c/(1+c))}.
\end{equation}
The value of $N_g(M)$ represents the number of photo-z galaxies inside
a halo with the mass of $M$ assuming that $N_g(M)$ is proportional
to $M$ as
\begin{equation}
N_{\rm g}(M)=\frac{M\rho_g(z)}{\rho_m(z)},
\end{equation}
where $\rho_m(z)$ and $\rho_g(z)$ are the average mass density and the
average number density of photo-z galaxies at redshift $z$.  Here the
halo mass is defined as $M_{\rm 200m}$
\begin{equation}
M\equiv\frac{4\pi r_{\rm 200m}^3}{3}\cdot 200\rho_m(z),
\end{equation}
where $r_{\rm 200m}$ is the radius corresponding to $M_{\rm 200m}$ and
then the concentration is defined as $c\equiv r_{\rm 200m}/r_s$.  We
adopt the mass-concentration relation from
\cite{DiemerKravtsov15}. The amplitude of the mass-concentration
relation is left as a free parameter -- i.e., the overall amplitude of
that relation is multiplied by a mass-independent factor $c_{\rm
  amp}$. We describe the projected cross-correlation function $w(R)$
by adding an off-centring term to model ``central'' galaxies that are
offset from halo centres as:
\begin{equation}
\label{eq:wR}
w(R)=\int \frac{k\,\mathrm{d}k}{2\pi} C_{\rm gg}(k)J_0(kR),
\label{eq:wR_app}
\end{equation}
and 
\begin{eqnarray}
C_{\rm gg}(k)&=&\frac{1}{\Delta\chi}\int \mathrm{d}M P(M)
\left(\frac{N_g(M)}{\rho_g(z)}\right)
\tilde{u}_{\rm NFW}(k;M) \nonumber \\
&& \times [q_{\rm cen}+(1-q_{\rm cen})\tilde{p}_{\rm off}(k;M)].
\end{eqnarray}
Here, we consider that the photo-z blurring just decreases the overall
amplitude of $w(R)$ (or the Fourier-transform $C_{gg}(k)$) without
changing the shape. The parameter $\Delta\chi$ represents the distance
corresponding to the photo-$z$ scatter. The function $\tilde{u}_{\rm
  NFW}(k;M)$ is the Fourier transform of the projected NFW profile
$u_{\rm NFW}(R;M)$ normalized with the mass $M$. The analytical
formula of $u_{\rm NFW}(R;M)$ is given by \cite{WrightBrainerd00}.
The parameter $q_{\rm cen}$ is the fraction of galaxies in the sample
that are located at the halo centre. The function
  $\tilde{p}_{\rm off}$ is the Fourier transform of the radial profile
  of off-centred galaxies, hereafter referred to as the off-centred
  profile. The radial profile of red galaxies around galaxies that are
  not central galaxies are flattened depending on the off-centred
  profile. We adopt two models to describe the off-centred profile of
  galaxies. One is the Gaussian offset model, i.e., $\tilde{p}_{\rm
    off}(k;M)=\exp(-k^2R_{\rm off}^2(M)/2)$
  \citep[e.g.,][]{OguriTakada11}.  The scale of this offset is defined
  as a fixed fraction of the halo radius as: $R_{\rm off}(M)=\tau_{\rm
    off}r_{\rm 200}(M)$ where $\tau_{\rm off}$ is left as a free
  parameter. The other is the NFW offset model, i.e., $\tilde{p}_{\rm
    off}(k;M)=\tilde{u}_{\rm NFW}(k;M)$ where the concentration
  parameter is proportional to the halo concentration as $c_{\rm
    off}c(M)$ and $c_{\rm off}$ is left as a free parameter. When
estimating the mean halo mass via Equation~\ref{eq:massdist}, we
integrate over the halo mass distribution of the sample, $P(M)$.
Since the redshift evolution for $P(M)$ is small during the current
redshift range as discussed in \cite{Simet17}, we do not include
redshift evolution and assume a single mean redshift of 0.25. To model
the cross-correlation functions, we limit our fits to
$0.1h^{-1}$Mpc$<R<2h^{-1}$Mpc, where contributions from the two-halo
term are negligible.

In addition, we also compare our off-centring models with
galaxy-galaxy lensing measurements.  The galaxy-galaxy lensing is
estimated using the excess surface mass density
\begin{equation}
\Delta\Sigma(R)\equiv \int \frac{k\, \mathrm{d}k}{2\pi} C_{\Sigma g}(k) J_2(kR),
\label{eq:dsigma}
\end{equation}
where $C_{\Sigma g}(k)$ is the galaxy-lensing cross-power spectrum
\begin{eqnarray}
C_{\Sigma g}(k)&=&\int \mathrm{d}M P(M) M \tilde{u}_{{\rm NFW}}(k;M) \nonumber \\
&& \times [q_{\rm cen}+(1-q_{\rm cen})\tilde{p}_{\rm off}(k;R_{\rm off}(M))]+
M_{\rm sub}. \nonumber \\
\end{eqnarray}
The last term represents the subhalo/baryonic core component
approximated as a stellar mass of $M_\ast$, which is set to be
$\log(M_\ast/h^{-1}M_\odot)=11.4$ \citep[e.g.,][]{Leauthaud16}.

\begin{table*}
\begin{center}
\begin{tabular}{ll}
  \hline\hline
Parameter & Description \\
\hline
$\Delta\chi [h^{-1}$Mpc] & Distance corresponding to photo-$z$ scatter  \\ 
$c_{\rm amp}$ & Amplitude of mass-concentration relation \\
$q_{\rm cen}^{\rm RMCG}$ & Fraction of central galaxies in RMCG sample \\
$q_{\rm cen}^{\rm BMEM}$ & Fraction of central galaxies in BMEM sample \\
$\tau_{\rm off}^{\rm High Pcen}$ & Ratio of off-centring scale to $r_{200}(M)$ in High Pcen sample 
in Gaussian offset model \\
$\tau_{\rm off}^{\rm RMCG}$ & Ratio of off-centring scale to $r_{200}(M)$ in RMCG sample 
in Gaussian offset model \\
$\tau_{\rm off}^{\rm BMEM}$ & Ratio of off-centring scale to $r_{200}(M)$ in BMEM sample
in Gaussian offset model \\
$c_{\rm off}^{\rm High Pcen}$ & Ratio of concentration parameter to $c(M)$ in High Pcen sample
in NFW offset model \\
$c_{\rm off}^{\rm RMCG}$ & Ratio of concentration parameter to $c(M)$ in RMCG sample
in NFW offset model \\
$c_{\rm off}^{\rm BMEM}$ & Ratio of concentration parameter to $c(M)$ in BMEM sample 
in NFW offset model \\
\hline
\end{tabular}
\end{center}
\caption{Free parameters in our modeling and their short descriptions.
\label{tab:params}}
\end{table*}

\begin{figure*}
\begin{center}
\includegraphics[width=16cm]{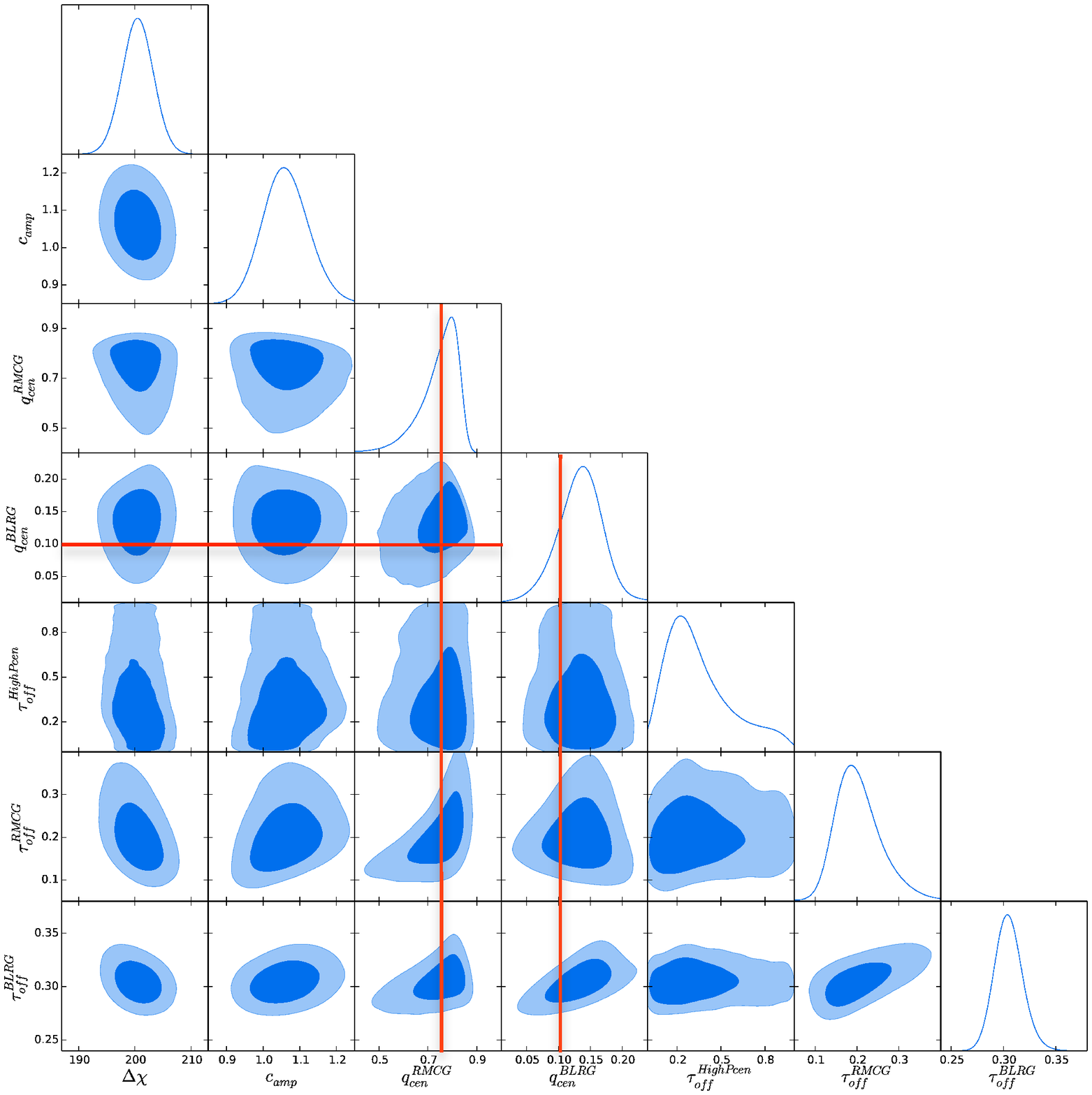}
\end{center}
\caption{1-$\sigma$ and 2-$\sigma$ constraints on the parameters
  obtained from the cross-correlation measurements $w(R)$ in the three
  samples. The distribution in the diagonal panels shows the posterior
  distribution of each parameter after marginalizing over the other
  parameters. The red solid lines represent the redMaPPer values of
  mean central fractions for the RMCG and BMEM samples. As shown, the
  model fits to cross-correlation data give consistent off-centring
  fractions with those inferred from the redMaPPer centring probabilities.
\label{fig:param_cont} 
}
\end{figure*}

\begin{table*}
\begin{center}
\begin{tabular}{cccccccc}
  \hline\hline
$\Delta\chi [h^{-1}{\rm Mpc}]$ & $c_{\rm amp}$ & $q_{\rm cen}^{\rm High Pcen}$ & $q_{\rm cen}^{\rm RMCG}$ & $q_{\rm cen}^{\rm BMEM}$ 
& $\tau_{\rm off}^{\rm High Pcen}$ & $\tau_{\rm off}^{\rm RMCG}$ & $\tau_{\rm off}^{\rm BMEM}$ \\
\hline
$200\pm 3 $ & $1.06\pm 0.06$ & $0.98\pm 0.01$ & $0.74\pm 0.10$ & $0.13\pm 0.04$ & $0.38\pm 0.25$ & $0.20\pm 0.06$ & $0.31\pm 0.01$ \\
\hline
\end{tabular}
\end{center}
\caption{1-$\sigma$ marginalized constraint on each parameter obtained
  by joint fitting of $w(R)$ for the High Pcen, redMaPPer central
  galaxy (RMCG), and the brightest member galaxy (BMEM) samples. The
  fitted parameters are 8 in total: the photo-z scatter $\Delta\chi$
  and concentration $c_{\rm amp}$, which are common for the three
  samples, and the central fraction $q_{\rm cen}$ and the relative
  offset $\tau_{\rm off}$ for three samples. We put the strong prior
  on the central fraction for High Pcen sample $q_{\rm cen}^{\rm High
    Pcen}=[0.95,1]$. The fitted range of scale from 0.1$h^{-1}$Mpc to
  2$h^{-1}$Mpc and the covariance among different samples in each bin
  in $R$ is taken into account.
\label{tab:offset}}
\end{table*}

\begin{table*}
\begin{center}
\begin{tabular}{cccccccc}
  \hline\hline
$\Delta\chi [h^{-1}{\rm Mpc}]$ & $c_{\rm amp}$ & $q_{\rm cen}^{\rm High Pcen}$ & $q_{\rm cen}^{\rm RMCG}$ & $q_{\rm cen}^{\rm BMEM}$ 
& $c_{\rm off}^{\rm High Pcen}$ & $c_{\rm off}^{\rm RMCG}$ & $c_{\rm off}^{\rm BMEM}$ \\
\hline
$200\pm 3 $ & $1.06\pm 0.06$ & $0.97\pm 0.01$ & $0.78\pm 0.09$ & $0.05\pm 0.05$ 
& $5.1\pm 2.8$ & $5.2\pm 2.8$ & $0.88\pm 0.20$ \\
\hline
\end{tabular}
\end{center}
\caption{Same as Table \ref{tab:offset}, but the results in the NFW
  offset model. The parameters related to off-centring scales
  $\tau_{\rm off}$ are replaced by $c_{\rm off}$.}
\label{tab:offset2}
\end{table*}

\section{Results}
We perform a simultaneous fit to the cross-correlation measurements of
the three samples (High Pcen, RMCG, and BMEM) using
Equation~\ref{eq:wR_app}.  There are 7 free parameters in our model
listed in Table \ref{tab:params}. Two of these parameters are common
for all these samples: the photo-z scatter $\Delta\chi$ and the
amplitude of the mass-concentration relation $c_{\rm amp}$.  There are
then five free parameters: the central fractions of RMCG and BMEM
samples $q_{\rm cen}^{\rm RMCG/BMEM}$ and the off-centring scales for
High Pcen, RMCG, and BMEM samples: off-centring scale relative to the
halo size $\tau_{\rm off}^{\rm High Pcen/RMCG/BMEM}$ in the Gaussian
offset model; concentration parameter in the off-centring
profile relative to the halo concentration parameter $c_{\rm off}^{\rm
  High Pcen/RMCG/BMEM}$ in the NFW offset model.

In this work, we assume that the High Pcen sample selects a high
fraction of central galaxies. For this sample, we therefore put a
strong prior on the central fraction $q_{\rm cen}^{\rm High
  Pcen}=[0.95,1]$, while the central fractions for other samples are
allowed to vary in the entire range $q_{\rm cen}^{\rm
  RMCG/BMEM}=[0,1]$.  We also add a loose prior $\tau_{\rm off}=[0,1]$
for all three samples and $c_{\rm off}=[0,10]$. Finally, we require
that the sum of the central fractions for RMCG and BMEM samples does
not exceed unity, that is, $q_{\rm cen}^{\rm RMCG}+q_{\rm cen}^{\rm
  BMEM}\le 1$.

Figure~\ref{fig:fit_angcor} shows the comparison of the observed
cross-correlation measurements for our best-fitting model. 
Our simple model using NFW profiles provides an excellent description
of the cross-correlation measurements.  The minimum chi-squared value
$\chi^2_{\rm min}$ is 16.3 for d.o.f$=37$ (45 data points minus 8
parameters) where the error is estimated by jackknife resampling and
the covariance among the samples at each bin of $R$ is included.  One
can see that the central fraction for the RMCG sample is much larger
than the BMEM sample, which is consistent with the values reported by
redMaPPer.

As a consistency check, we also compare the galaxy-galaxy lensing
measurements $\Delta\Sigma$ with the prediction of the theoretical
model (Eq.~\ref{eq:dsigma}).  We use the best-fitting values of
$q_{\rm cen}$ and $\tau_{\rm off}$ obtained from the cross-correlation
measurements.  We do not use
$\Delta\chi$ and $c_{\rm amp}$ parameters as they are only relevant
for the clustering cross-correlation measurement and not for the
underlying matter distribution probed by the cluster-galaxy lensing.
As shown in Figure~\ref{fig:fit_dsigma}, we find that the lensing
measurements are consistent with our theoretical model. The $\chi^2$
values are 27.5 for the 24 data points, while for the High Pcen
sample, the $\chi^2$ value is 7.2 for 8 data points.

Figure~\ref{fig:param_cont} shows the constraints on our parameters
from the cross-correlation measurements $w_{gg}$.  Red solid lines
show the mean values reported by redMaPPer for $(q_{\rm cen}^{\rm
  RMCG}, q_{\rm cen}^{\rm BMEM})$. These are consistent with our
constraint obtained from the cross-correlation measurements within
$1\sigma$. Tables~\ref{tab:offset} and \ref{tab:offset2} list the
marginalized constraints on each parameter. Our constraint on the
central fraction is $0.74\pm 0.09$ for RMCG samples and $0.13\pm 0.04$
for BMEM samples, consistent with the redMaPPer mean values of 0.754
and 0.103, respectively. The average values of $r_{200}(M)$ are
1.1$h^{-1}$Mpc for the three samples.  The physical off-centring scale
converted from the fitted $\tau_{\rm off}$ values is
0.2-0.3$h^{-1}$Mpc consistently across the three samples. The error on
the off-centring scale for ``High Pcen'' is large since the
off-centring fraction is so low.  When adopting the NFW offset model,
the constraints on the central fraction becomes $0.77\pm0.09$ for RMCG
samples and $0.05\pm 0.05$ for BMEM samples, which is also consistent
with the redMaPPer estimates.

We note that our results rely on the prior that the High Pcen sample
has a nearly true radial profile of photo-z red galaxies without
off-centring, i.e., $q_{\rm cen}^{\rm High Pcen}=[0.95,1]$. Since the
galaxies in the High Pcen sample usually have well-defined central
galaxies, the prior is reasonable. The posterior distribution of
$q_{\rm cen}^{\rm High Pcen}$ however has a peak at 0.95, the minimum
edge of our prior. This may indicate that our modeling does not
perfectly describe the observed galaxy radial profile for several
reasons (e.g., baryonic feedback effect). Nevertheless, our results
for the off-centring values are determined by the difference ratio of
the radial profiles among the samples.  As long as the High Pcen
sample has a true galaxy radial profile, the off-centring values for
the other samples are barely affected by the details in the modeling
of the radial profile.

\section{Summary}
We have investigated whether photometrically derived redMaPPer
centring probabilities are valid for the cluster sample in which the
brightest member galaxy (BMEM) is not the redMaPPer central galaxy
(RMCG), that is, the galaxy with the highest centring probability in
each cluster.  We measure the projected cross correlation function of
each cluster 'centre' with photometric red galaxies, which is
sensitive to the off-centring fraction. We use ``High Pcen'' sample
where the RM centring probabilities is larger than 99\% as a reference
sample of central galaxies.  We find that the centring fraction is
$74\pm 10$\% for RMCGs and $13\pm 4$\% for BMEM when the radial
profile of off-centred galaxies follows a Gaussian form. When adopting
the NFW model for the off-centred profile, the constraints on the
centring fraction becomes $78\pm 9$\% for RMCGs and $5\pm 5$\% for
BMEM. The values are consistent with the redMaPPer values, 75\% for
RMCG and 10\% for BMEM. Our analysis is a strong self-consistency test
of the RM centring probabilities. Our results indicate that the
redMaPPer centroids are better tracers of the centre of the cluster
potential than the brightest cluster members, and that the redMaPPer
algorithm provides accurate estimates of centring probabilities for
the proposed cluster centres.

We also measure galaxy-galaxy lensing to find that the measurements
are consistent with our lensing models with the best-fitting
off-centring values obtained from the cross-correlation
measurements. The statistical error is much larger than that of the
cross-correlation measurements because of the large shot noise due to
the limited number of source galaxies. The lensing signal will be
substantially improved by using upcoming deeper imaging surveys such
as Subaru Hyper-Suprime Cam and Dark Energy Survey, which have a much
higher number density of source galaxies than SDSS, enough to
compensate for their smaller areas.

We show that the redMaPPer centring algorithm provides reliable
estimates of centring probabilities using the cross-correlation
measurements. This is consistent with the previous work by
\cite{RozoRykoff14} who studied the centring probability by
comparison with X-ray clusters. Our result supports the work by
\cite{Hoshino15} who investigated the central occupation of LRGs and
BMEMs based on the redMaPPer centring probabilities. They found that
20$\sim$30\% of redMaPPer clusters the brightest cluster member
galaxies are not central galaxies. The redMaPPer centring algorithm
is also useful for calibrating off-centring effect in various studies
such as mass estimates using stacked lensing analysis
\citep{Johnston07,Leauthaud10,Okabe10,OguriTakada11,George12,Oguri17}.  RSD
studies using reconstructed ``halo'' catalogs have the uncertainty of
the off-centring fraction \citep{Hikage12}, which can be mitigated
using the redMaPPer centring values.

\section*{Acknowledgments}
CH is supported by MEXT/JSPS KAKENHI Grant Numbers 16K17684. RM is supported by the Department of
Energy Early Career Award program.

\vspace{1cm}

\bibliographystyle{mn2e} \bibliography{mn-jour,ref} \label{lastpage}

\end{document}